# Unexpected aspects of strain relaxation and compensation in InGaAs metamorphic structures grown by MOVPE

*Agnieszka M. Gocalinska\*,[†] Marina Manganaro[†,§] and Emanuele Pelucchi[†]*

[†] Tyndall National Institute, "Lee Maltings", University College Cork, Cork, Ireland

**ABSTRACT** We present a selection of stack designs for MOVPE grown $In_xGa_{1-x}As$ metamorphic buffer layers following various convex-down compositional continuous gradients of the In content, showing that defect generation and strain can be managed in a variety of ways, some rather unexpected (and unreported). Indeed, we observe that it is possible to grow surprisingly thick tensile strained layers on metamorphic substrates, without significant relaxation and defect generation. We believe our findings give significant insights to the investigation of strain, relaxation and defect distribution in metamorphic buffer design, so to obtain properly engineered/tailored structures (the most successful ones already finding applications in device growth).

**Introduction**

Tuneability is a critical feature in custom-design of semiconductor optical devices. The range of wavelengths provided by compositional change in III-V alloys can be broadened by variable quantum confinement.[1] However, some wavelengths are hard to achieve, as the requirements for sufficient external barrier providing confinement and the quantum well (QW) composition might demand large lattice mismatch between the two, which may lead to strain-induced defect formation. Also, passive devices like solar cells and detectors have much higher efficiency when the absorbing material is thicker, calling for a "bulk" active layer, likely to relax when not perfectly lattice-matched to the substrate. The very limited generic semiconductor substrate range is therefore a large obstacle.

One largely explored approach to overcome this problem is to adjust the substrate lattice constant by step- or gradual change of the buffer composition[2,3,4,5] to reach the desired in-plane lattice parameter. The design of such metamorphic buffer layer (MBL) needs to meet certain requirements, especially for dislocation density[6] (i.e., as low as possible), but also to allow cost-effective growth, i.e. the thickness should be kept to a minimum. Therefore, whilst optimizing the designs in general, one needs to have in mind the highest possible lattice parameter change in minimal thickness, while preserving high surface quality (i.e. minimum appearing surface dislocation density, which often corresponds to minimum roughness).

Step-graded buffers with just a few highly mismatched steps usually do not bring the best possible structural results, as the rapid relaxation leads to high defectivity[7]. Gradually graded buffers, either continuous or dense multi-step ones, usually allow for higher control over defect distribution[8]. The preferable epilayer relaxation is then obtained by creation of misfit dislocations, whose density corresponds to the compositional grading rate (lattice mismatch per thickness unit) and should be kept below a (specific) critical value, having a detrimental effect on the surface topography. Ideally, in a proper design, individual dislocations are given the possibility to glide for relatively long distances, providing the most efficient degree of strain relaxation.[9,10]

Sublinearly graded layers were found to have lower average misfit dislocation density and less residual in-plane strain.[11] Choosing a specific curve allows for partially pseudomorphic growth with low defect density, preferentially in the top part of the layer, allowing for a surface organization resulting in "good" quality surface3. This unfortunately often has the drawback of leaving significant residual strain in the structure: e.g. the final part of the layer is compressively strained and the in-plane lattice parameter is smaller than the "natural", relaxed lattice parameter of the material of a given composition. Therefore, whilst designing such metamorphic buffers, it is usually necessary to bring the grading to higher compositions than the nominal value, so to obtain the desired in-plane lattice constant.[12]

We have recently demonstrated how, with the use of a specific growth design, it is possible to breach the gap between GaAs and InP[13] or InP and InAs[14] with a modified parabolic-gradient-based In-Ga exchange curves. It should be said that, while the adjustment of the lattice parameter has been reported to give better morphological results when the composition of the alloy follows an exchange of group V, not III, elements[15], $In_xGa_{1-x}As$ has been historically the material of choice, partially because the competition between group V elements on the surface during epitaxial growth, which makes the effective composition harder to control.

Here, we discuss in some detail specific engineered structures allowing to obtain intermediate lattice parameters, some relevant conditions for stacking different designs, as well as we report an unexpected capacity of the system to accommodate large strains, permitting for "bulk" growth of highly mismatch, good quality layers. The relevance of strain compensating techniques is also discussed in this context. Examples of such designed MBLs already served as buffer layers in successful device growths[13,16], and are being developed further for implementation in other applications. Moreover, our particular interest is the use of GaAs-based MBL is the production of semiconductor lasers (i.e. we need to allow for thick structures) operating at the technologically important 1.3 µm and 1.55 µm wavelengths.[17] The metamorphic growth allows for reaching a lattice parameter intermediate between GaAs and InP and therefore provides opportunity for band structure engineering, design and optimization of a semiconductor laser with broader degrees of freedom available than in the case of a binary compound substrate (i.e. higher confinement and larger set of dielectric constants, potentially leading to significantly improved performances compared to InP-based devices).[18]

**Methodology**

All epitaxial samples discussed here were grown in a high purity MOVPE[19,20,21] commercial horizontal reactor (AIX 200) at low pressure (80 mbar) with purified $N_2$ as carrier gas. The precursors were trimethylindium (TMIn), trimethylgallium (TMGa), arsine ($AsH_3$) and phosphine ($PH_3$). The samples' designs were varied, as described in the text. Graded buffers started from GaAs and were initiated with minimal controllable In flow, therefore the real initial composition can be estimated to be between 0.00 and 0.01 In. All samples had a homoepitaxial GaAs 200 nm thick buffer grown prior to the graded $In_xGa_{1-x}As$. Growth conditions were: V/III ratio 130, growth rate 1 µm/h, estimated real growth temperature 650 ºC; samples were grown on perfectly oriented (100) ± 0.02º GaAs substrates (unless stated otherwise when particular sample is described). All layers were nominally undoped.

All epitaxial growths described herein as successful, resulted in smooth surfaces with cross-hatch pattern clearly visible when inspected with an optical microscope in (Nomarski) Differential Interference Contrast (N-DIC) or in dark field mode (not shown). Subsequent detailed morphological studies were performed with Atomic Force Microscopy (AFM) in tapping/non-contact mode at room temperature and in air. The defects formation and propagation were observed with cross sectional Transmission Electron Microscopy (TEM) under bright field condition, with 200 keV accelerating voltage, lenses: C1 = 2000 µm and C2 = 100 µm, objective aperture = 40 µm to maximize contrast. TEM images were recorded in [110] orientation. The defect density was estimated by a recently demonstrated method based on large scale, multiple AFM imaging supported by TEM.[22] The accuracy of the method has constraints, and we refer the interested reader to Ref. 22.

The assessment of composition and the strain in the layers was made according to measurements of Reciprocal Space Maps (RSM) obtained by high resolution X-ray diffraction measurements (HRXRD). Measurements were done in a symmetric (004) and two asymmetric (224 and -2-24) reflections with sample positioned at 0 °, 90 °, 180 ° and 270 ° with respect to its main crystallographic axes (the calculations followed Vegard's law, which is a standard method for calculating alloy composition and strain in partially relaxed III-V materials, see e.g., Ref. 23, 24, 25, 26). Details regarding all discussed samples are summarized in Table 1.

**Table 1** Growth conditions, design and characterisation of all discussed MBLs.

| Figure | Sample | Substrate misorientation | Growth temperature | MBL nominal structure | RMS* [nm] | composition[†] In [fraction] | in-plane lattice parameter[†] [Å] | "relaxed" composition[†] In [fraction] | residual parallel strain[†] $\varepsilon_p$ [%] / (relaxation in respect to GaAs) | Estimated AFM/TEM defect density [×] [cm$^{-2}$] |
|---|---|---|---|---|---|---|---|---|---|---|
| **1** | A | (100) ± 0.02° | 650 °C | 1000nm parabolic In$_x$Ga$_{1-x}$As grading ~0<x<0.33 | 3 | 0.330 | 5.7612 | 0.2665 | -0.0044 (80.77%) | <5·10$^5$ |
| **2a** | B | (100) + 2° tow. [1-10] ± 0.1° | 660 °C | 200nm In$_{0.12}$Ga$_{0.88}$As  200nm In$_{0.25}$Ga$_{0.75}$As  500nm In$_{0.4}$Ga$_{0.6}$As | 10 | 0.398 | 5.7999 | 0.3621 | -0.0025 (90.99%) | too defected to estimate |
| **2b** | C | (100) ± 0.02° | 650 °C | 800nm In$_x$Ga$_{1-x}$As parabolic grading ~0<x<0.23  800nm In$_x$Ga$_{1-x}$As parabolic grading 0.18<x<0.41  300nm In$_{0.41}$Ga$_{0.59}$As | 7 | 0.425 | 5.7907 | 0.3391 | -0.0060 (79.72%) | 5·10$^6$ |
| **2c** | D | (100) ± 0.02° | 650 °C | 800nm In$_x$Ga$_{1-x}$As parabolic grading 0<x<0.23  180 nm In$_{0.1}$Ga$_{0.9}$As  800nm In$_x$Ga$_{1-x}$As parabolic grading 0.18<x<0.41  300nm In$_{0.41}$Ga$_{0.59}$As | 4 | 0.408 | 5.7941 | 0.3476 | -0.0042 (85.21%) | 2·10$^6$ |
| **2d/e** | E | (100) ± 0.02° | 650 °C | 800nm In$_x$Ga$_{1-x}$As parabolic grading ~0<x<0.23  180 nm In$_{0.1}$Ga$_{0.9}$As  800nm In$_x$Ga$_{1-x}$As parabolic grading 0.18<x<0.41  125nm In$_{0.3}$Ga$_{0.7}$As | 5 | 0.301 | 5.7751 | 0.3101 | +0.0006 (97.23%) | <5·10$^5$ |

| | | | | | | | | | | | |
|---|---|---|---|---|---|---|---|---|---|---|---|
| 3 | F | (100) ± 0.02° | 650 °C | 1000nm $In_xGa_{1-x}As$ parabolic grading ~0<x<0.33 | | 2 | 0.160 | 5.7445 | 0.2253 | +0.0046 (70.88%) | <5·10$^5$ |
| | | | | 700nm $In_{0.16}Ga_{0.84}As$ | | | | | | | |
| 4a/b | G | (100) ± 0.02° | 650 °C | 800nm $In_xGa_{1-x}As$ parabolic grading ~0<x<0.23 | | 12 | 0.514 | 5.7836 | 0.4274 | -0.0060 (83.23%) | 10$^8$ |
| | | | | 180 nm $In_{0.1}Ga_{0.9}As$ 800nm $In_xGa_{1-x}As$ parabolic grading 0.18<x<0.41 | | | | | | | |
| | | | | 260 nm $In_{0.26}Ga_{0.74}As$ 800nm $In_xGa_{1-x}As$ parabolic grading 0.31<x<0.53 | | | | | | | |
| 4c/d | H | (100) ± 0.02° | 650 °C | 1000nm $In_xGa_{1-x}As$ parabolic grading ~0<x<0.33 | | 15 | 0.476 | 5.8123 | 0.3927 | -0.0058 (82.52%) | too defected to estimate |
| | | | | 1140nm $In_{0.16}Ga_{0.84}As$ 1000nm $In_xGa_{1-x}As$ parabolic grading 0.23<x<0.53 | | | | | | | |

\* Calculated from 10x10 [μm$^2$] AFM images after standardised flattening

† Values corresponding to final grown layer, estimated by XRD measurements

˟ Values corresponding to final grown layer, estimated by combination of cross sectional TEM and top-view AFM, as discussed in Ref. 22.

**Results and discussion**

In Figure 1 we present a representative/reference characterization of an MBL, 1 μm thick (Sample A). The $In_xGa_{1-x}As$ (~0<x<0.33) layer was grown following the previously reported design of single parabolic exchange curve3. It shows a "flat" (~3 nm RMS), step-bunched surface and relatively low defect density (estimated $<5 \cdot 10^5$ cm$^{-2}$) towards the end of the layer. Most of the defects are buried down close to the GaAs substrate, as it is clear from TEM in Figure 1c). The in-plane lattice parameter in this growth is equivalent to a fully relaxed $In_{0.27}Ga_{0.73}As$, as estimated by XRD measurements. The TEM image is actually in agreement with that: the thickness corresponding to the end of the defected region (~550 nm from the bottom of the growth) can be translated into approximately the same indium composition value (~0.27), i.e. the growth proceeds from there on pseudomorphically.

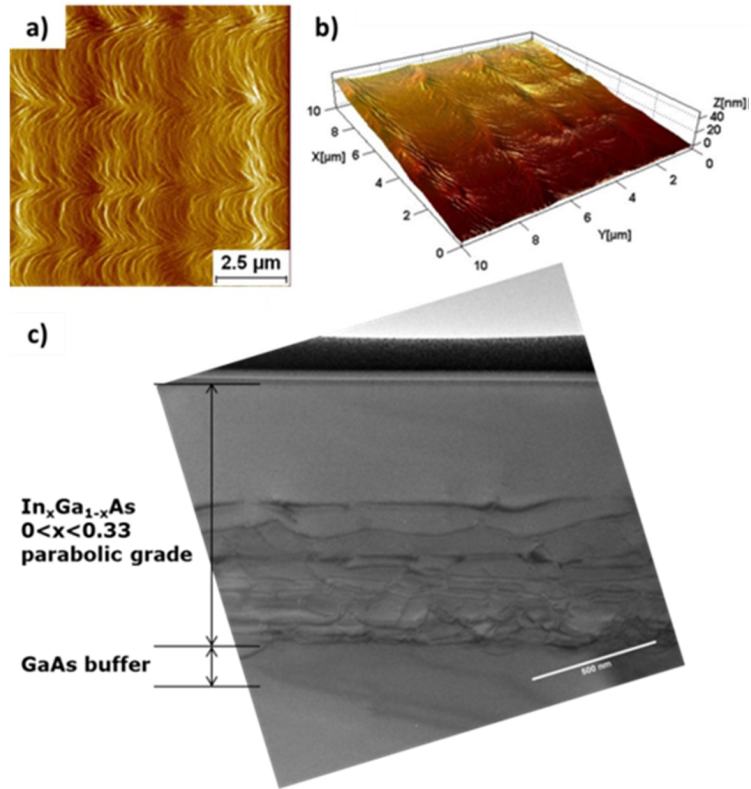

**Figure 1 (color online). Surface morphology and defect distribution in MBL following single parabolic exchange curve (Sample A): AFM images (a) signal amplitude, b) reconstructed 3D height image, and c) cross-section TEM in [110] orientation**

These results as such are not unexpected in view of the existing literature (see e.g. 3, 27). Accordingly, we have also successfully grown samples with lower final In content with excellent surface quality. Following a similar design we could increase In content as far as around 0.40 (maintaining the overall thickness to 1 μm) with no detectable defects in cross-sectional TEM nor top view AFM (therefore with estimated defect density below $5 \cdot 10^5$ cm$^{-2}$, which is the detection limit for our methodology, see Ref. 22). Nevertheless the top morphology was showing already a significant roughening, probably linked to the build-up of residual strain (not shown). Increasing indium composition even further resulted in morphological and eventually relaxation problems. Therefore we would estimate, in a 1 μm thick structure, that a reasonable limit for the maximum final indium composition in growths following a single parabolic In grading to be between 0.33-0.40, in our growth conditions with this design.

To reach higher In compositions, maintaining similar surface quality, we could either increase the layer thickness along with the final indium value or introduce modifications to the exchange curve. The first option is not ideal, as the large portion of the growth following single parabolic grading design is pseudomorphic – therefore there is a high residual strain affecting the morphology and the in-plane lattice parameter is not necessarily increasing. One possible option to avoid this is to redesign the In-Ga exchange curve. After the highest defect density region we switched the rate to follow along a tangent to the parabola, extending the moderate defect density part. The growth was finish of by reducing the curve slope towards the end of the grading to achieve a low defect density ending. We have reported on the successful implementation of that elsewhere[13,14].

Here, within other things, we'd like to discuss an alternative approach: stacking of several gradings on top of each other. This solution benefits from releasing the majority of the strain in high defect density regions while allowing for full surface reconstruction towards the end of each block. The added advantage is permitting for, to some extent, (residual) strain management. The drawback is an unavoidable increasing thickness of the stack. We point out that hereafter, the consecutive grading was always initiated at the In composition equal to the one corresponding to in-plane lattice parameter of underlying buffer (we call this "step back" in this manuscript).

As a first step in our discussion we present on Figure 2 (see Table 1 for details) a comparison of several different growth designs with the same final layer composition. Growing two parabolic gradings on top of each other, implementing strain managing by step back (Figure 2 b), Sample C) brought a result (RMS = 7 nm) which is comparable with the simple step graded buffer of lower thickness (Figure 2 a), Sample B: RMS of 10 nm). Inserting a low In composition layer (In=0.10, 180 nm) in between the gradings, which adds a tensile strained region between them (strain balancing layer, SBL), improved the overall morphology significantly (Figure 2 c and Supplementary Figure 1, Sample D) lowering the RMS to 4 nm. It should be stressed that, apart from the SBL, both Samples C and D were grown with identical design and under same growth conditions.

To facilitate a precise evaluation of the composition and strain by XRD we have overgrown these samples with a (thick) cap layer with a composition corresponding to the end of the grading (a practice we utilized also elsewhere in the development process and common in the community, see e.g. Ref. 8). Nevertheless, this adds a significant amount of elastic energy accumulated in the structure, as the cap is highly strained, and has potentially detrimental effect on the surface morphology. Replacing it with a cap of a composition lattice-matched to the in-plane lattice parameter (Figure 2 d) and e), Sample E) shows further improvement of the morphology (RMS = 5 nm). The reduced thickness of the cap (125 nm in Sample E in comparison to previously used 300 nm final composition layer in Samples C and D) should not affect the final roughness estimation, as from our experience the lattice-matched overgrowth is neither flattening nor roughening the surface in the discussed materials and structures (in this thickness range).

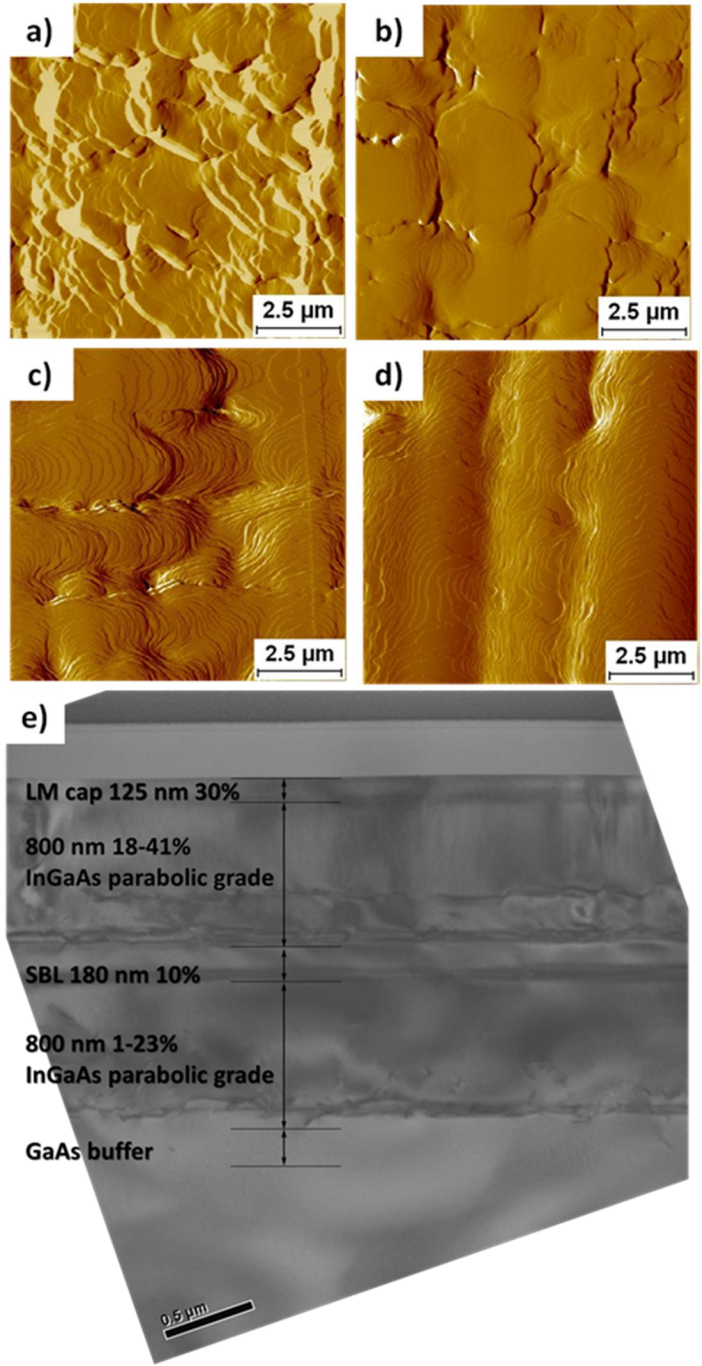

**Figure 2 (color online) AFM images (signal amplitudes) of a) step graded buffer (Sample B); b) double parabolic grading (Sample C), c) double parabolic grading with SBL (Sample D); d) double parabolic grading with SBL and lattice-matched cap (Sample E); e) cross-section TEM in [110] orientation of the latter. Exact designs of samples are listed in Table 1.**

Moreover, the surface quality improved also in another aspect, showing a lower overall defectivity: perpendicular lines indicating threading dislocations are not present (as measured with AFM).[22] In our interpretation and methodology, it suggests we reduced the estimated defect density by an order of magnitude (to below $5·10^5$ cm$^{-2}$ for Samples E (and F actually, as discussed later on), from initial mid-range $10^6$ cm$^{-2}$ in Samples C and D).

The TEM of Sample E is presented on Figure 2 e) and it shows an expected defect distribution: high defect density at the initial part of the grading and pseudomorphic growth towards the end of parabola, repeated exactly in the second parabolic stage.

Since our further discussion will concentrate on the unexpected phenomenology of the SBL, one point needs clarifying. The composition and thickness of the SBL discussed above were initially estimated based on the simplifying assumptions that the pseudomorphic part of the MBL if fully strained and the defected part fully relaxed. Therefore the accumulated elastic energy $E$ can be calculated from the following formula:

$$E_{MBL} = \int_{h_3}^{h_2} \varepsilon_p^2(h) \cdot Y(y) dh$$

where: $h$ – thickness, $Y$ – Young's modulus, $y$ - indium concentration, $\varepsilon_p$ - in-plane strain (variables with indices $_2$ and $_3$ refer to respective values at the beginning and the end of the strained (pseudomorphic) region, as indicated on graph on Figure 3 a)). Since in the case of a continuous parabolic grading, the composition $y$ relates to thickness $h$ by the following3:

$$y(h) = (y_3 - y_0) \cdot \left(1 - \left(1 - \frac{h}{h_3}\right)^2\right) + y_0$$

To balance the accumulated elastic energy fully, a tensile SBL needs to have $E_{SBL} = -E_{MBL}$, where

$$E_{SBL} = \varepsilon_p^2 \cdot Y(y) \cdot h$$

for a constant composition layer, and the minus has been added to show the opposite strain contribution.

An SBL needs to fulfil additional criteria: the lattice parameter offset between the end of previous layer and the SBL cannot exceed the critical value leading to creation of dislocations (or of strain-induced dot like structures), but it needs to be large enough to provide good interfacial stress for dislocation glide. Also small thicknesses would be preferable, in general, with the scope of maintaining the overall thickness the minimum possible. The above discussed rough model brought positive results, (Sample E), with a significant reduction of surface roughness as we have already shown in Figure 2 d) and Table I. However, interestingly and unexpectedly, the experimental verification of our initial assumption that the SBL should balance out the fully strained part of the grading turned out to be an over simplistic picture (see Supporting Information for exact calculations for samples E and F).

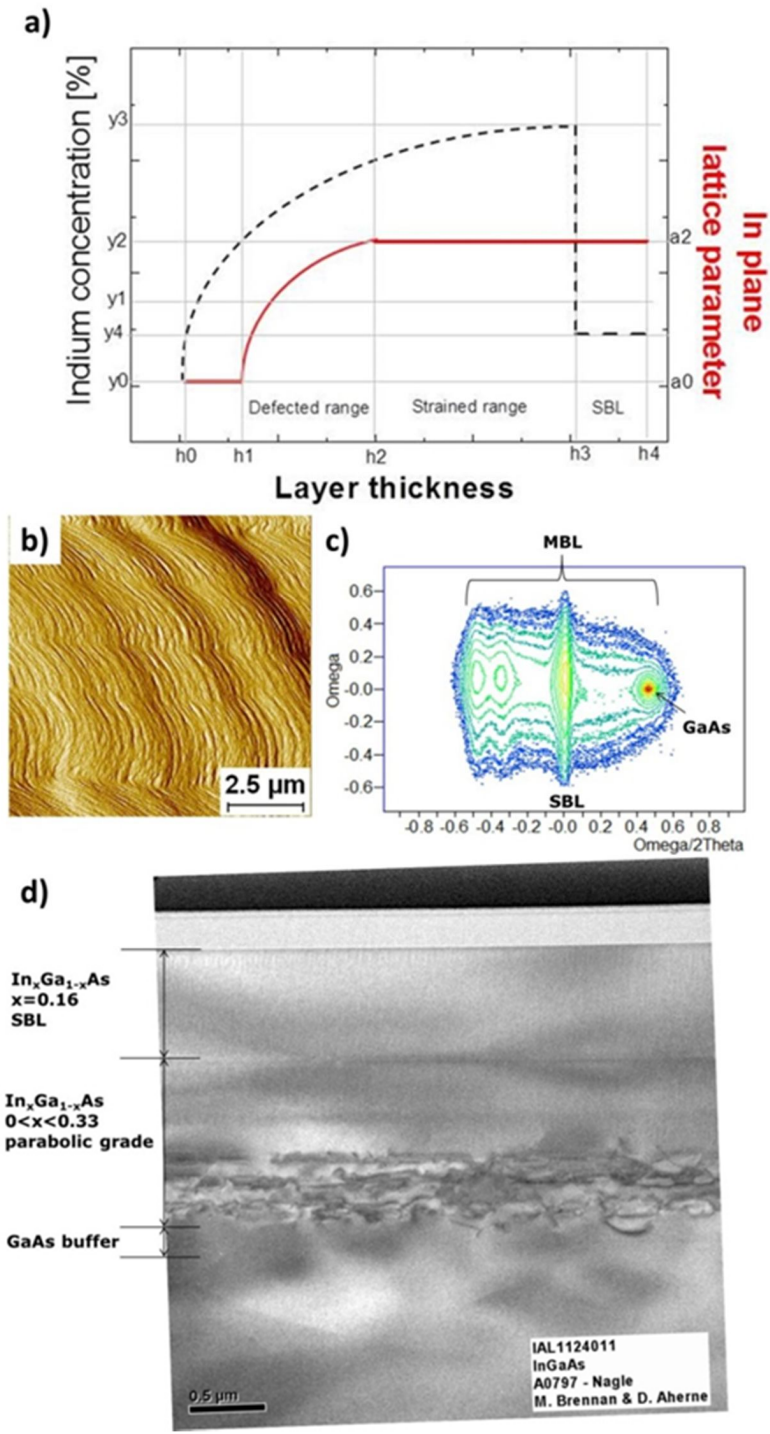

**Figure 3 (color online) Single parabolically graded MBL with SBL (Sample F): a) simplified design sketch of single parabolic grading with SBL (black dash curve: alloy composition regarding indium concentration, red curve: in-plane lattice parameter change in the structure – before the relaxation threshold the substrate lattice parameter is preserved, then increases during the defected part and finally settles at a value preserved for the rest of the structure in the pseudomorphic fraction of MBL and in the SBL); b) AFM image (signal amplitude) of the top surface; c) 2-axis X-ray diffractogram in 004 reflection; d) cross-section TEM in [110] orientation.**

Indeed, we were able to grow layers on top of the gradings with thickness exceeding several times the estimated critical value without any signs of relaxation due to tensile strain. In Figure 3 we present a simple example of an SBL grown on parabolic grading (Sample F): with end-of-grading composition In=0.33, in plane residual strain -0.0044% and a pseudomorphic region thickness of 450 nm (calculated based on TEM of grading-only, as from Figure 1 c)). We should have been able to balance the compressive strain by implementing ~50 nm thick, 0.16 In layer. Nevertheless, as it can be clearly seen from the images in Figure 3d) (Sample F, and Figure c), XRD), even 700 nm thick SBL could be grown without any signs of relaxation, i.e. we exceeded the compensating thickness by 650 nm without any significant defect generation (we observe that a full relaxation would have resulted in a highly visible network of dislocations, clearly appearing in any, single, TEM image). A surprising result if one thinks that a reasonable estimate for the critical thickness for a layer grown on another with a difference in indium composition of only 0.04 is around 100-150 nm or so (obviously, this is just an approximation, as all depends on the specific exact conditions and In composition). Indeed, here the difference in lattice parameter is significantly higher (roughly 0.10 considering the estimated in plane parameter at the end of the first grading), i.e. one would expect a strong relaxation.

The strained pseudomorphic nature of the SBL is also confirmed in the following (Sample F): the 2 axis diffractogram presented on Figure 3 c) shows a typical pattern for a parabolic metamorphic structure: from the GaAs peak position on the right, the MBL signal stretches out towards higher lattice parameters, with signal intensity increasing (with lattice parameter) as, by design, the material corresponding to higher indium concentration is thicker (In concentration increases slower in later parts of the MBL). Despite "locking" the in-plane lattice parameter at roughly half way through the grading, the diffracted signal is still being observed further to the left, as XRD measures the perpendicular out-of-plane lattice parameter. The thick SBL layer grown on top has a lower In content than the in-plane lattice parameter of the grading and additionally is tensile strained, therefore shows it's signal at significantly lower lattice constants (in the middle of the grading). Also, the lack of significant broadening of that peak in Omega/2Theta direction suggests the layer is pseudomorphic (despite its significant thickness) in respect to the end of the grading, confirming the previous analysis by cross-sectional TEM (Figure 3 d)).

We note that the defected region of the MBL in this specific case is thinner than in the first growth discussed (Sample A, where it was 550 nm): from TEM we could estimate that in this last sample (Sample F) it spans across ~450 nm. Also the XRD estimated in-plane strain of the end of the grading and SBL is higher, limiting the in plane lattice parameter to ~$In_{0.23}Ga_{0.73}As$. While those are however small variations, that can be assigned most likely to run to run reproducibility and uncertainty of the TEM and XRD measurements, we note that nevertheless this might also indicate changes of the overall structural relaxation process.

Indeed, the highly tensile strained region was, obviously, much thicker than it would have been predicted from theoretical calculation for strained $In_xGa_{1-x}As$.[28] While we also do not observe any *significant* signs of phase separation (see e.g. 2, 29 and references therein) in the tensile strained region (Figure 3 d), we caution the reader on the limits of the technique in this respect, which here would simply show a modulation in the TEM contrast/colour. In our case the phase separation could have been partially suppressed by the high temperature and V/III ratio used during the growth, which are known to have that effect in MOVPE[30]. We also observe that our AFM analysis[22] suggests a very small number of defects (the methodology of Ref. 22, based on counting defect related features by AFM, should be taken carefully in its capability of giving absolute, exact values, nevertheless the relative reduction in defect number is likely to be reliable). The relaxation by plastic deformation of the layer would be expected at a much lower layer thickness, as we discussed. This might suggest a different mechanism of relaxation in this system, e.g. by increasing the defect density in the partially relaxed buried part of the MBL or creation of an extra crystallographic tilt[31,32]. Anyway, the choice of the parabolic Ga-In exchange curve to confine the defects close to the interface and the possibility to (over)compensate the built-up compressive strain in a full stack make these very useful tools for design engineering. The interchanging of compressive and tensile strained layers should/could also promote the bending of many of the remaining threading dislocations at interfaces, provided that the heteroepitaxial stress is high enough. We would also like to note that a similar behaviour was reported in Ref. 33, where tensile strained silicon was grown on a virtual SiGe substrates. The authors observed a less effective lattice relaxation (silicon layers grown on virtual substrates were shown to exceed many times the critical thickness and exhibit limited levels of relaxation), and they associated this with the tensile state of silicon, with a major role attributed to the appearance of microtwins. In the lack of extensive microscopy analysis, we are not assuming a similar mechanism in our case, but nevertheless this might be a possible direction for discussion and new investigations.

Nevertheless there are (interesting) limitations for the extent for the implementation of these techniques. In Figure 4 (and Table 1) we present two alternative growths designed to reach $In_{0.53}Ga_{0.47}As$ in three (Figure 4 a) and b)) and two step (Figure 4 c) and d)) MBLs, both featuring the strain balancing technique. In both cases after reaching ~0.45 In (as estimated by in-situ monitoring[14] (not shown), with good agreement with TEM) a new generation of defects starts to appear, regardless of the growth design and strain compensation applied. Those "new" defects (and associated possible phase separations) are not visibly initiated at the highly defected parabolic regions nor are they threading from the bottom of the growth. Here the overall accumulated elastic energy of the system is also fully or over-balanced, as e.g. the Sample H presented in Figure 4 c) had an SBL over 1 µm thick with large lattice constant offset, exceeding several times the value estimated for full strain compensation. Yet Sample H turned out to be defective on large scale, as indicated by AFM (Figure 4 d); RMS = 15 nm) and showed a significant broadening of the XRD signal (not shown). This is probably related to segregation issues in high In-containing alloys and local strain. If so, it needs to be handled by other means, for example by surfactant exploitation or with specific substrate offcut selection, as we discussed in Ref. 13. Nevertheless, the examples shown in Figure 4 are anyway a good verification that strain engineering allows for obtaining control over defect distribution – the highly defected regions are each time confined to the beginning of the parabolic gradings and are followed by the pseudomorphic part, till the growth reaches the critical In content.

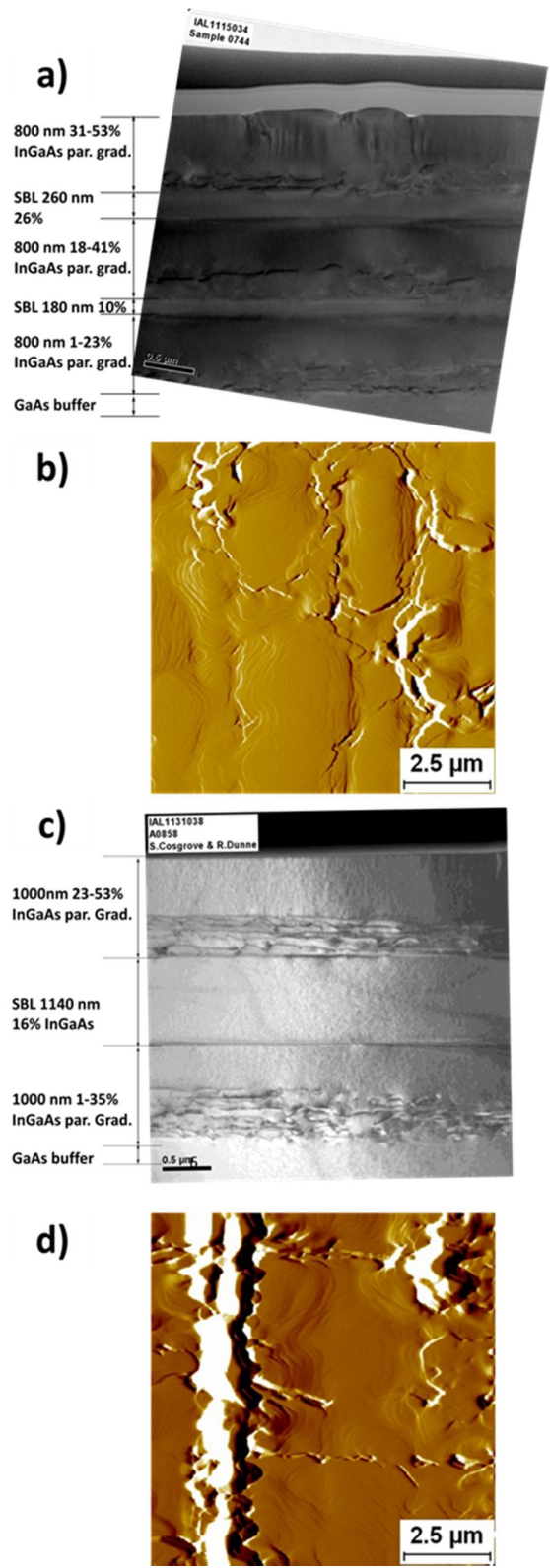

**Figure 4** Examples of multistep stacks, showing defect distribution and material limitations : a) cross-section TEM in [110] orientation and b) AFM (signal amplitude) of triple parabolic grading with 2 SBLs (Sample G); c) cross-section TEM in [110] orientation and d) AFM (signal amplitude) of double parabolic grading with SBL (Sample H).

**Conclusions**

We have discussed several possible growth designs oriented on changing the original lattice parameter of the substrate to a desired value involving strain management in implementing $In_xGa_{1-x}As$ metamorphic buffer layers. We discussed the possibility of alternating continuously graded buffers as well as more complex structures featuring strain balancing layers: constant composition barriers inserted between the gradings, with low In concentration, counteracting the compressive strain in pseudomorphic parts of the growth to reduce the overall elastic energy accumulated in the structure. We surprisingly discovered that very thick, significantly tensile strained layers can be grown on top of an MBL, with good surface quality and no detectable defects in cross-sectional TEM and/or AFM topography. Nevertheless, limitations to this approach for the $In_xGa_{1-x}As$ gradings were found, in respect to the design of the exchange curve and maximum In composition.

## ASSOCIATED CONTENT

**Supporting Information**

Exemplary calculation of composition and thickness of strain balancing layer grown on a metamorphic buffer layer.

Figure presenting cross-section TEM of Sample D.

## AUTHOR INFORMATION

**Corresponding Author**

* e-mail agnieszka.gocalinska@tyndall.ie, Phone: +353 21 420 6625

**Present Addresses**

[§] Inst. de Astrofísica de Canarias, E-38200 La Laguna, Tenerife, Spain; Universidad de La Laguna, Dpto. Astrofísica, E-38206 La Laguna, Tenerife, Spain

**Author Contributions**

The manuscript was written through contributions of all authors. All authors have given approval to the final version of the manuscript. AG designed, grew and characterized the samples. MM had supporting role in growth and characterization. EP supervised and coordinated the works. AG and EP wrote the manuscript together.

**Funding Sources**

Ireland under grants 10/IN.1/I3000, 12/RC/2276 (IPIC) and by an EU project FP7-ICT under grant 258033 (MODE-GAP).

**ACKNOWLEDGMENT**

We gratefully acknowledge Dr. K. Thomas for the MOVPE system support, Intel Corporation for partial financial support and Intel Ireland for TEM analysis (S. Cosgrove, Y. Khalifa, C. Williams, R. Dunne and D. Aherne) and other support (R. Nagle).

For Table of Contents Use Only

Manuscript title: Unexpected aspects of strain relaxation and compensation in InGaAs metamorphic structures grown by MOVPE

Authors: Agnieszka M. Gocalinska, Marina Manganaro, and Emanuele Pelucchi

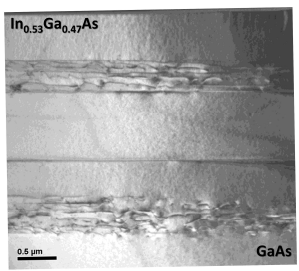

Synopsis: TEM of a complex design of an $In_xGa_{1-x}As$ metamorphic buffer layer